\documentclass[epjST]{svjour}
\usepackage[utf8]{inputenc}
\usepackage{graphicx}
\usepackage{flafter}
\usepackage{amsmath,amssymb,amstext}
\usepackage{braket}
\usepackage{float}
\usepackage{color}
\newcommand{\bk}[1]{\left(#1\right)}					
\newcommand{\Bk}[1]{\left[#1\right]}					

\newcommand{\bvec}[1]{\boldsymbol{#1}}					
\newcommand{\absq}[1]{\left|#1\right|^2}				
\begin{document}
\title{Time-periodic Klein tunneling through optomechanical Dirac barriers }
\author{Christian Wurl\thanks{\email{wurl@physik.uni-greifswald.de}} \and Holger Fehske\thanks{\email{fehske@physik.uni-greifswald.de}} }
\institute{Institute of Physics, Ernst Moritz Arndt University Greifswald, Greifswald, 17489, Germany}

\abstract{
We study the interconversion between photons and phonons coupled via radiation pressure in artificial Dirac materials realized by a honeycomb array of optomechanical cells. In particular we analyze the chiral tunneling of (photon-phonon) polaritons  through an oscillating planar barrier. While a static barrier accommodates constructively interfering optical or mechanical waves leading to photon or phonon transmission, an oscillating barrier allows for inelastic scattering that causes sideband excitations and interference effects which, in turn, may suppress or revive the light-sound interconversion.
} 
\maketitle

{\it Introduction.} In optomechanical graphene, that is, a honeycomb array of optomechanical cells  driven by a laser with frequency $\omega_{l}$, co-localized cavity photon (eigenfrequency $\omega_{o}$) and phonon (eigenfrequency $\omega_{m}$) modes interact (linearly) via radiation pressure. The latter is tunable by the laser power. Recently, the scattering and conversion process between photons and phonons, triggered by static laser-induced planar~\cite{SPM15} and circular quantum barriers~\cite{WF17}, has been worked out within an effective Dirac-Weyl theory, and the emergence of optomechanical Dirac physics has been demonstrated.  Because of the chiral nature of the quasiparticles, having a  Dirac-like bandstructure, the transport phenomena show similarities to those of low-energy electrons in graphene,  but are more subtle due the photon-phonon coupling in the barrier, leading to the formation of polariton (photon-phonon) states. Moreover, for perpendicular incidence of the photon wave, Klein tunneling appears, that is, the unimpeded transmission of the particle regardless of the height or width of the barrier. Interestingly, in the limit of low photon energies or high coupling strengths, when the barrier acts as a kind of Fabry P\'{e}rot interferometer, a perfect interconversion between photons and phonons takes place, as a result of a constructive interference of standing optical and mechanical waves respectively.

In this contribution, we extend these investigations by analyzing the passage  of Dirac-Weyl quasiparticles in optomechanical graphene through a harmonically oscillating (driven) potential barrier, i.e., we consider the significant case where the energy is not conserved.  To this end, we solve the time-periodic scattering problem for a perpendicularly impinging plane photon wave of energy $E$ (injected by a probe laser), and discuss how a quantum barrier that oscillates in time with frequency $\Omega$ affects the tunneling process. Since Klein tunneling persists for oscillating barriers due to the conservation of helicity, we expect that the transport through the barrier is mainly determined by the conversion rate between photons and phonons.

{\it Theoretical approach.} 
For sufficiently low energies, if the barrier is smooth on the lattice scale but sharp on the de Broglie wavelength, umklapp scattering is suppressed and the continuum approximation applies. Then, the system under consideration can be described by the  optomechanical Dirac-Weyl  Hamiltonian
\begin{equation} \label{H}
H=\bk{\overline{v}+\frac{1}{2}\delta v \, \tau_{z}}\bvec{\sigma} \cdot \bvec{k}-g\bk{x,t}\tau_{x}\,, 
\end{equation}
given in units of $\hbar$ after rescaling $H\rightarrow H-\hbar\omega_{m}$~\cite{SPM15,WF17}. Here, $\overline{v}=(v_{o}+v_{m})/2$, $\delta v=v_{o}-v_{m}$, with $v_{o,m}$ as the Fermi velocities of the optical/mechanical modes, $\bvec{\tau}$ and $\bvec{\sigma}$ are Pauli matrices, $\bvec{k}$ ($\bvec{r}$) gives the wavevector (position vector) of the Dirac wave, and $g\bk{x,t}$ parametrizes the time-dependent photon-phonon coupling.  Note that the single-valley Dirac-Hamiltonian~\eqref{H} is obtained within the rotating-wave approximation, i.e., in a frame rotating with the laser frequency, in the red detuned moderate-driving regime, $\Delta=\omega_{l}-\omega_{o}=-\omega_{m}$.  In order to make the scattering inelastic, we assume that the laser amplitude is modulated by a frequency much smaller than the eigenfrequency of the laser and the mechanical mode,  $\Omega \ll \omega_{l}, \omega_{m}$ (otherwise the rotating-wave approximation is not granted). Furthermore, to stay within the continuum approximation,  $\Omega$ should be much smaller than the typical mechanical hopping in the array,  i.e. $\Omega \ll 2v_{m}/3a$ with $a$ as the lattice constant~\cite{SPM15}. Then the coupling strength within the barrier of width $w$ is given by 
\begin{equation} \label{potential}
g\bk{x,t}=\Bk{g_{0}+g_{1}\cos \bk{\Omega t}}\Bk{\Theta\bk{x}-\Theta\bk{x-w}}\,,
\end{equation}
where $g_{0,1}$ is assumed to be constant and $g_{1}\leq g_{0}$.

For the tunneling problem, we consider the incoming photon to be in a plane wave state at energy $E$ and use the time-independent eigensolutions of~\eqref{H} for $x<0$, $\psi^{in}\sim\ket{o}\text{exp}(ik^{o}x-iEt)$, with $k^{o,m}=E/v_{o,m}$  being the optical/mechanical wavenumber and $\ket{o,m}$ the bare optical/mechanical eigenstate of $\tau_{z}$. In case of  perpendicular photon incidence and a barrier potential that is translational invariant in  $y$-direction, the scattering problem becomes one-dimensional. Then, the helicity, $\sigma_{x} \cdot k_{x}/|k_{x}|$, is a conserved quantity with eigenvalue $+1$ (this quantum number is therefore omitted in the following). For this reason no reflected waves appear and Klein-tunneling takes place. As a result of the optomechanical coupling, behind the barrier $x>w$, the transmitted wave  consists of optical and mechanical modes $\psi^{t}=\psi^{t;o}+\psi^{t;m}$, $\psi^{t;o,m}\sim\sum_{n}t^{o,m}_{n}\ket{o,m}\text{exp}(ik_{n}^{o,m}x-iE_{n}t)$. Here, energy states with $E_{n}=E+n\Omega$ and $k^{o,m}_{n}=E_{n}/v_{o,m}$, $n\in \mathbb{Z}$, are superimposed, since the oscillating barrier can give (take) energy to (away from) photons and phonons. The wave inside the barrier is $\psi^{b}=\psi^{b;+}+\psi^{b;-}$, $\psi^{b;\pm}\sim \sum_{n,n'}b_{n}^{\pm}[c_{nn'}^{o;\pm}\ket{o}+c_{nn'}^{m;\pm}\ket{m}]\text{exp}(iq_{n}^{\pm}x-iE_{n-n'}t)$, with Fourier coefficients $c_{nn'}^{o,m;\pm}$ and wavenumbers $q_{n}^{\pm}$ obtained from Floquet theory~\cite{SHF15_2,S65}. It matches with the incident and transmitted wavefunction at the boundaries, which defines an infinite system of coupled linear equations for the scattering coefficients $t_{n}^{o,m}$ and $b_{n}^{\pm}$. From its numerical solution we obtain the current density of the transmitted wave: 
\begin{equation}
j^{t;o,m}\bk{x/v_{o,m}-t}=v_{o,m}\sum _{n,n'}(t^{o,m}_{n'})^{*}t^{o,m}_{n}  \text{exp}\Bk{i\bk{n-n'}\Omega\bk{x/v_{o,m}- t}}.
\end{equation}
Then the equation of continuity gives the time-averaged transmission probability $\overline{T}^{o/m}=(v_{o,m}/v_{o}) \sum_{n}\absq{t^{o,m}_{n}}$, with  $\overline{T}^{o}+\overline{T}^{m}=1$ (no backscattering). Because there are no phonon waves impinging on the barrier,  the transmission probability $\overline{T}_{m}$ can be understood as photon-phonon conversion probability.

{\it Results.} In what follows, we adopt $v_{m}=0.1v_{o}$ and employ units such that $v_{o}=1$. Moreover, since the scattering problem is invariant under the transformation $[E,g_{0,1},\Omega,w^{-1}]\rightarrow  \gamma [E,g_{0,1},\Omega,w^{-1}]$ with $\gamma \in \mathbb{R}$, we use units such that $\Omega=1$.

For a static barrier, $g_{1}=0$, we can analytically calculate the transmission probability  of the mechanical mode,
\begin{equation} \label{transprob}
T^{m}_{st}=[1+(k^{o})^2 v_{o}^2\bk{v_{o}-v_{m}}^2/\bk{4v_{m}v_{o}g_{0}^2}]^{-1}\sin ^2 \bk{(q^{+}_{st}-q^{-}_{st})w/2},
\end{equation}
with wavenumbers $q^{\pm}_{st}$ obtained from the static energy dispersion~\cite{WF17}. 
\begin{figure}
\center
\includegraphics[width=0.45\textwidth]{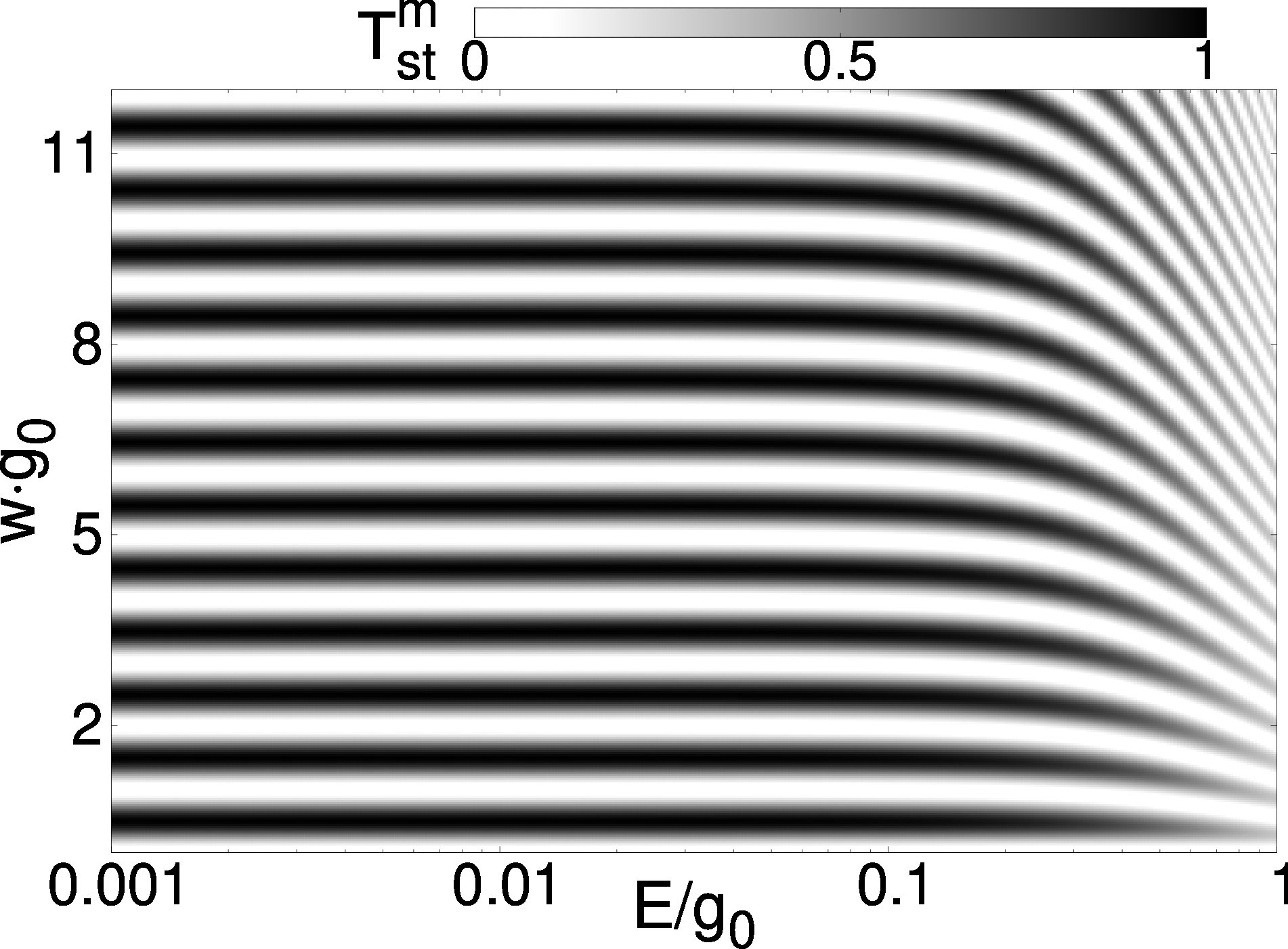}
\hspace{0.3cm}
\includegraphics[width=0.45\textwidth]{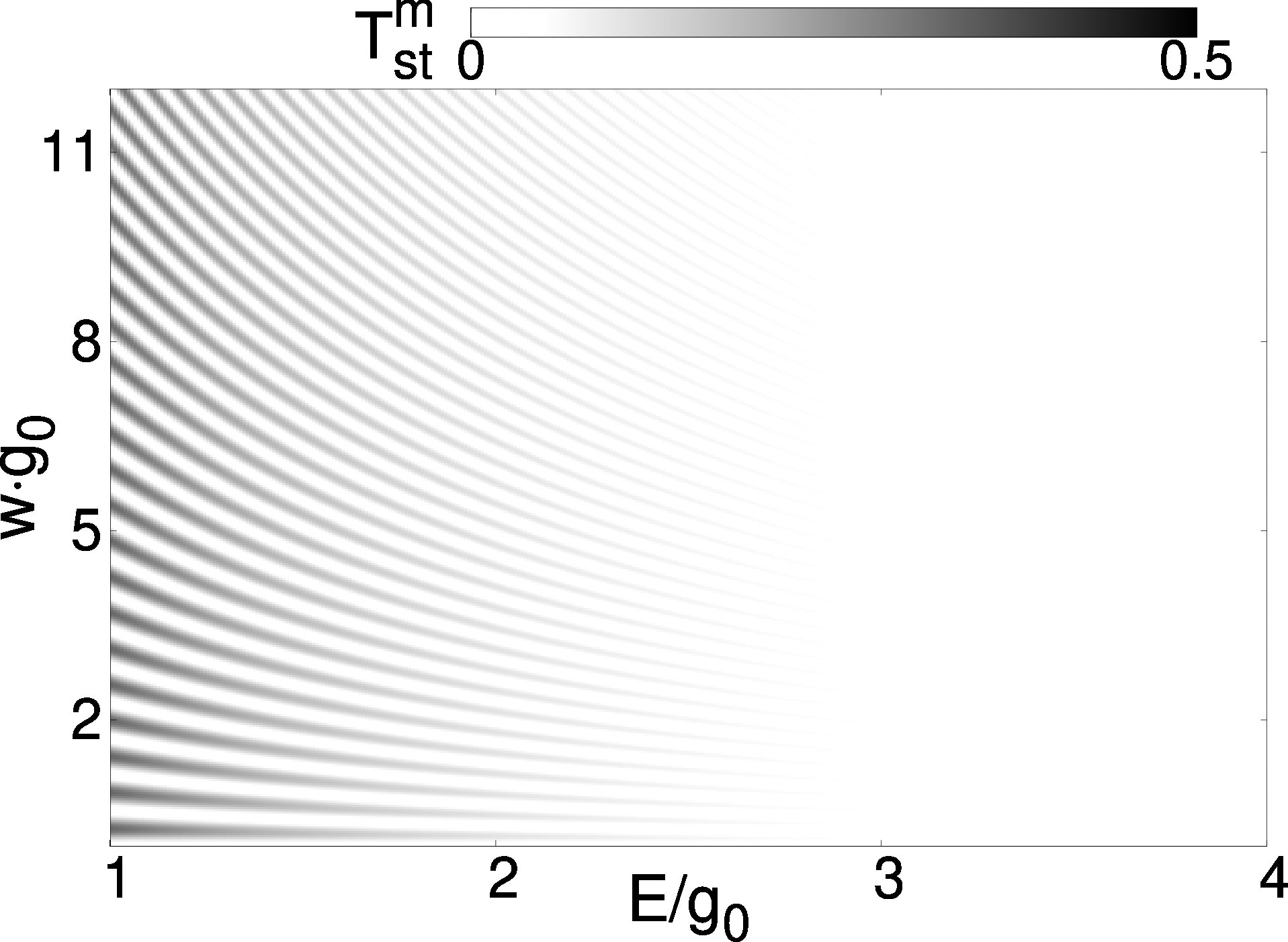}
\includegraphics[width=0.45\textwidth]{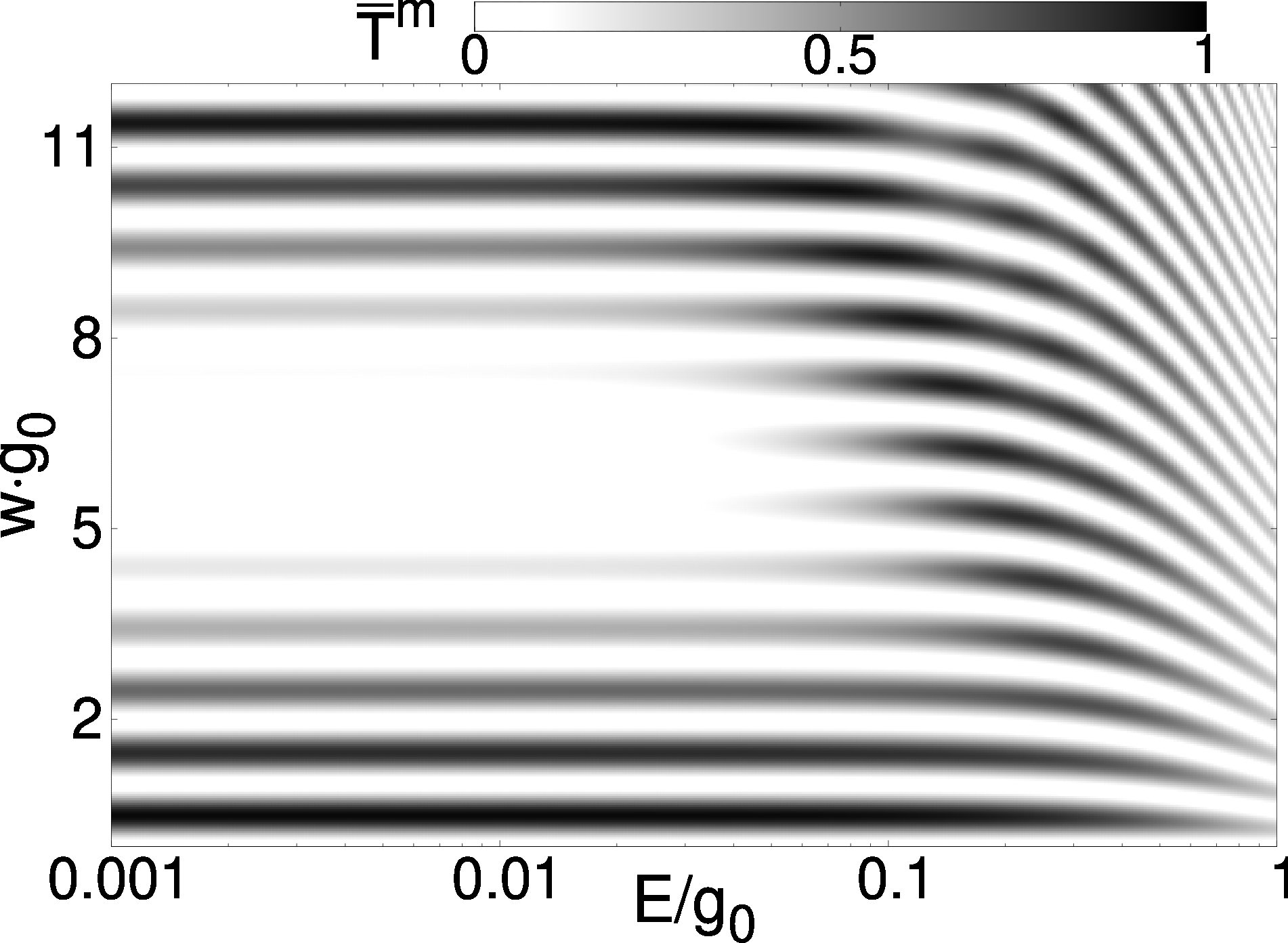}
\hspace{0.3cm}
\includegraphics[width=0.45\textwidth]{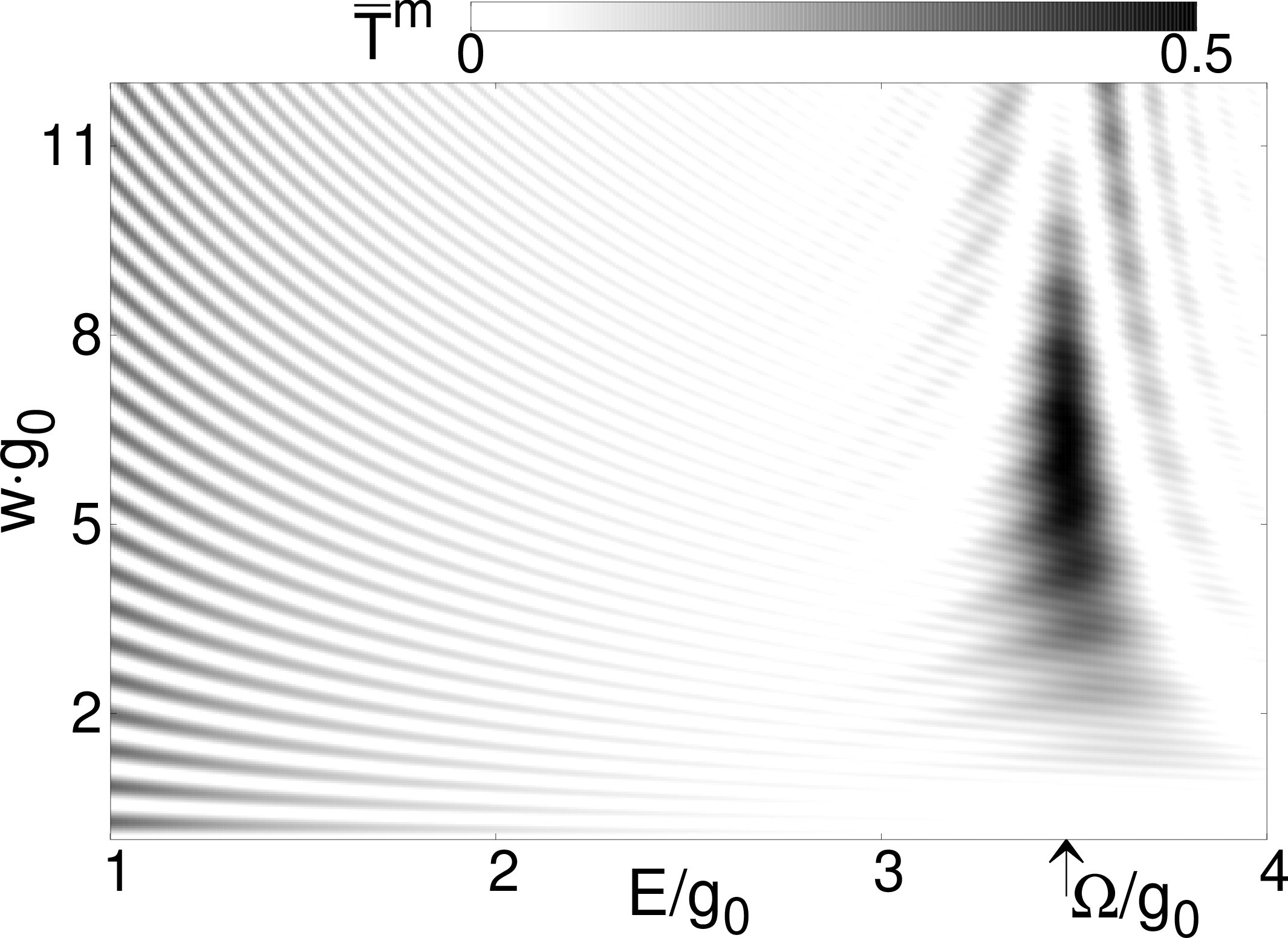}
\caption{Transmission probability of the phonon in the $E/g_{0}$-$wg_{0}$ plane for a static barrier ($g_{1}=0$; top panels) and for an oscillating barrier ($g_{1}=0.073\Omega$, $g_{0}=0.287\Omega$; $\Omega/g_{0}\approx 3.48$; bottom panels). The transmission probability of the photon is $T_{st}^{o}=1-T_{st}^{m}$ ($\overline{T}^{o}=1-\overline{T}^{m}$).}\label{fig1}
\end{figure}
\noindent
\begin{figure}
\center
\includegraphics[width=0.85\textwidth]{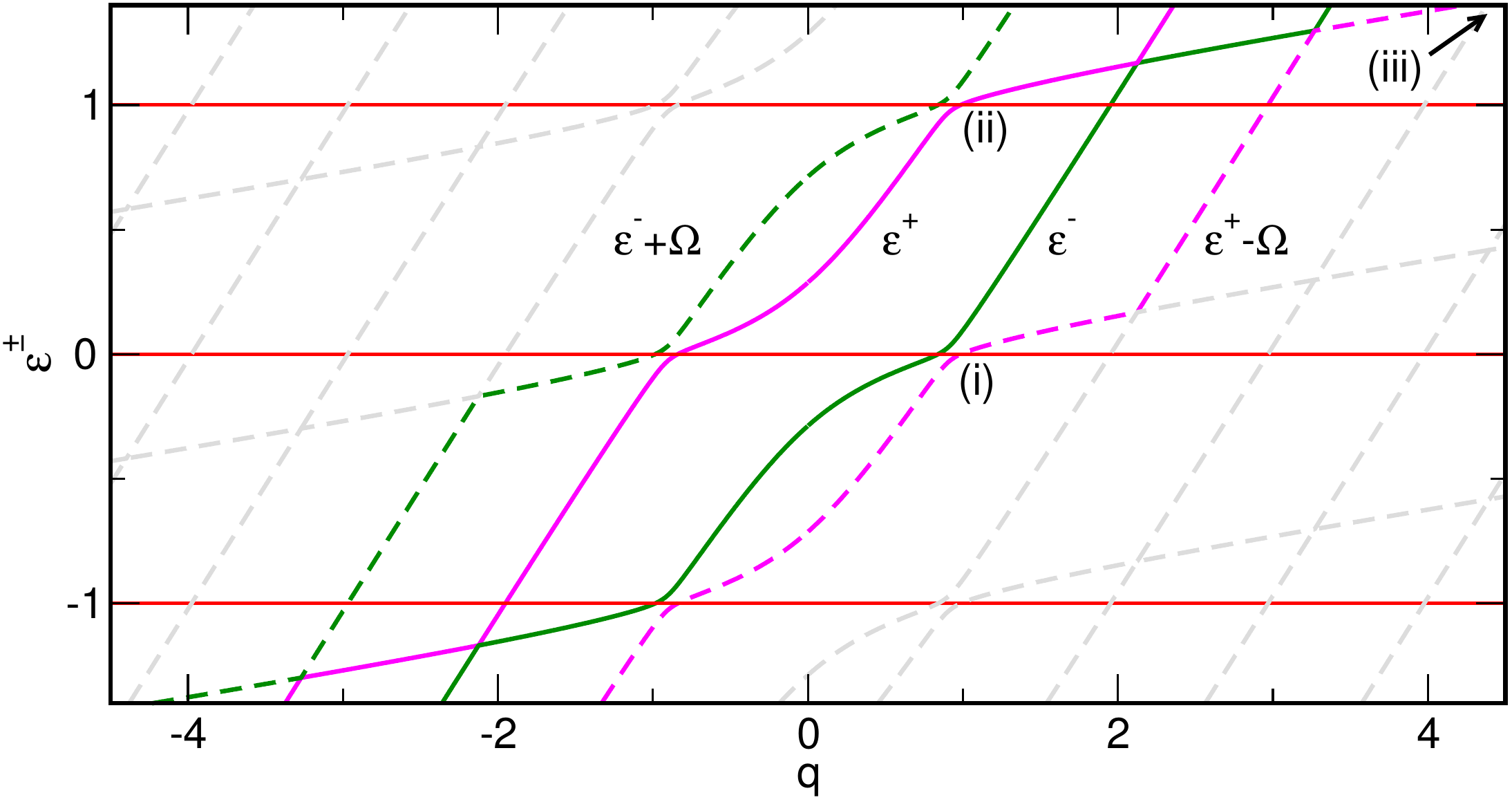}
\caption{(Color online) Floquet quasienergies $\varepsilon^{\pm}$  as a function of wavenumber $q$ for the parameter values of Fig.~\ref{fig1}, lower panels. The quasienergies are defined in such a way that they coincide with the energy dispersion of the static case  for $q \rightarrow 0$. Avoided crossings occur between $\varepsilon^{\pm}$ and $\varepsilon^{\mp}\pm \Omega$ (further quasienergy bands are marked in grey).}\label{fig2}
\end{figure}
Since $T_{st}^{o}=1-T_{st}^{m}$, in Fig.~\ref{fig1} (upper panels) only $T^{m}_{st}$ is plotted in the $E/g_{0}$-$wg_{0}$ plane. As a result of the optomechanical coupling, the incoming photon can be  converted into phonons, i.e., $T^{m}_{st}>0$. For energies larger than the barrier height (right panel), $T^{m}_{st}$ reveals a stripe structure with low intensity, where for  $E/g_{0}\gtrsim 2$ the photon-phonon conversion is strongly suppressed since $v_{m}\ll v_{o}$~\cite{WF17}. For energies smaller than the barrier height (left panel), the stripes in $T^{m}_{st}$  are much more pronounced, especially in the limit of small energies (high coupling strengths) $E/g_{0}\rightarrow 0$. Then the two polaritonic waves inside the barrier have antiparallel wavenumbers $\pm g_{0}/\sqrt{v_{o}v_{m}}=\pm q_{st}$ and interfere in such a way, that  the wave function simplifies to $\psi^{b}\propto \cos (q_{st}x) g_{0} \ket{o}+i v_{o}q_{st} \sin (q_{st}x)\ket{m}$. In this way, in a semiclassical perspective, the barrier acts as a kind of Fabry-Pérot interferometer accomodating  standing optical and mechanical waves respectively. If the optical (mechanical) wave interferes constructively with itself, the transmission becomes purely photonic (phononic), $T^{m}_{st}=0$ ($T^{m}_{st}=1$), where the resonance condition is $wg_{0} =\sqrt{v_{o}v_{m}} n \pi/2 \simeq 0.5 n $ with $n$ even (odd) natural number.

\begin{figure}
\center
\includegraphics[width=0.97\textwidth]{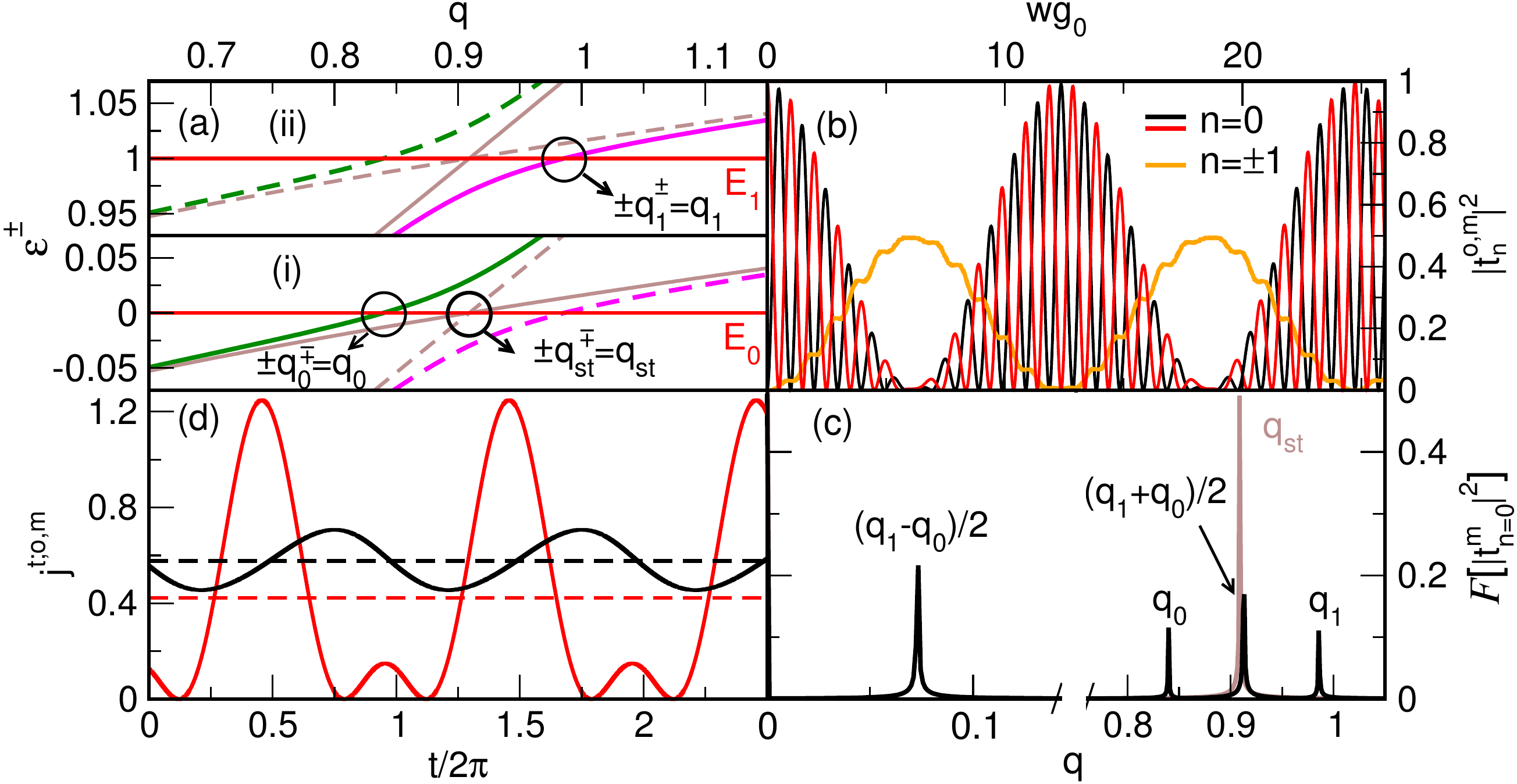}
\caption{(Color online) Scattering characteristics for photon energy $E \simeq 0$.  (a) Quasienergies $\varepsilon^{\pm}$ as a function of  $q$ for the cases (i) and (ii) of Fig.~\ref{fig2}.  Drawn in are the energies of the central band $E_{0}$ and the first sideband $E_{1}$ (red lines). Marked are wavenumbers $q=q_{0,1}$, for which $\varepsilon^{\pm}(q)=E_{0,1}$. For comparison the two polariton branches of the dispersion of the static case, $E^{\mp}(q)$ (solid)  and $E^{\pm}(q)\mp\Omega$ (dashed), are shown (brown) with wavenumber $q_{st}$. Avoided crossings appear in the vicinity of points, where the two static polariton branches cross each other.  (b) Transmission probability for optical/mechanical central bands $n=0$ (red/black) and first excited optical sideband $n=\pm 1$ (orange) as a function of  $wg_{0}$ (here, $|t_{n}^{m}|^2$ is multiplied by $v_{m}/v_{o}$).    (c) Fourier spectrum of $F[|t^{m}_{n=0}|^{2}](q)$ [for comparison the Fourier spectrum of $T^{m}_{st}$ is included (brown line)]. (d)  Time evolution of the optical/mechanical  current density $j^{t;o,m}$ (red/black) and the corresponding time-averaged current density (dashed)  at $x=w$ for $wg_{0}= 2.346$ [crossing of red and orange lines (b)]. In all panels, $g_{0}=0.287\Omega$, $g_{1}=0.073\Omega$ (corresponding to $E/g_{0}=10^{-3}$ in Fig.~\ref{fig1}).}\label{fig3}
\end{figure}

An oscillating barrier may cause inelastic scattering by excitation of states with energies shifted by  multiples of the oscillation frequency, $E_{n}=E+n\Omega$. In addition to the central band, sidebands now yield a significant contribution to the transmission of polaritons. The number of sidebands involved in the tunneling process is mainly determined by the ratio $g_{1}/\Omega$. Even for weak couplings $g_{1}$ (high oscillation frequencies $\Omega$), i.e., in the antiadiabatic limit $g_{1} / \Omega \ll 1$, the transmission pattern of $T_{st}^{m}$ is modified by a few sidebands only, see Fig.~\ref{fig1} (lower panels). At very low photon energies, $E/g_{0} \ll 1$, for certain widths of the barrier the transmission becomes purely photonic. By contrast, at photon energies close to  resonance, $E \sim \Omega$ (in Fig.~\ref{fig1} at $E/g_{0} \approx 3.48$), the transmission of phonons may be  significantly increased. 

\begin{figure}
\center
\includegraphics[width=0.97\textwidth]{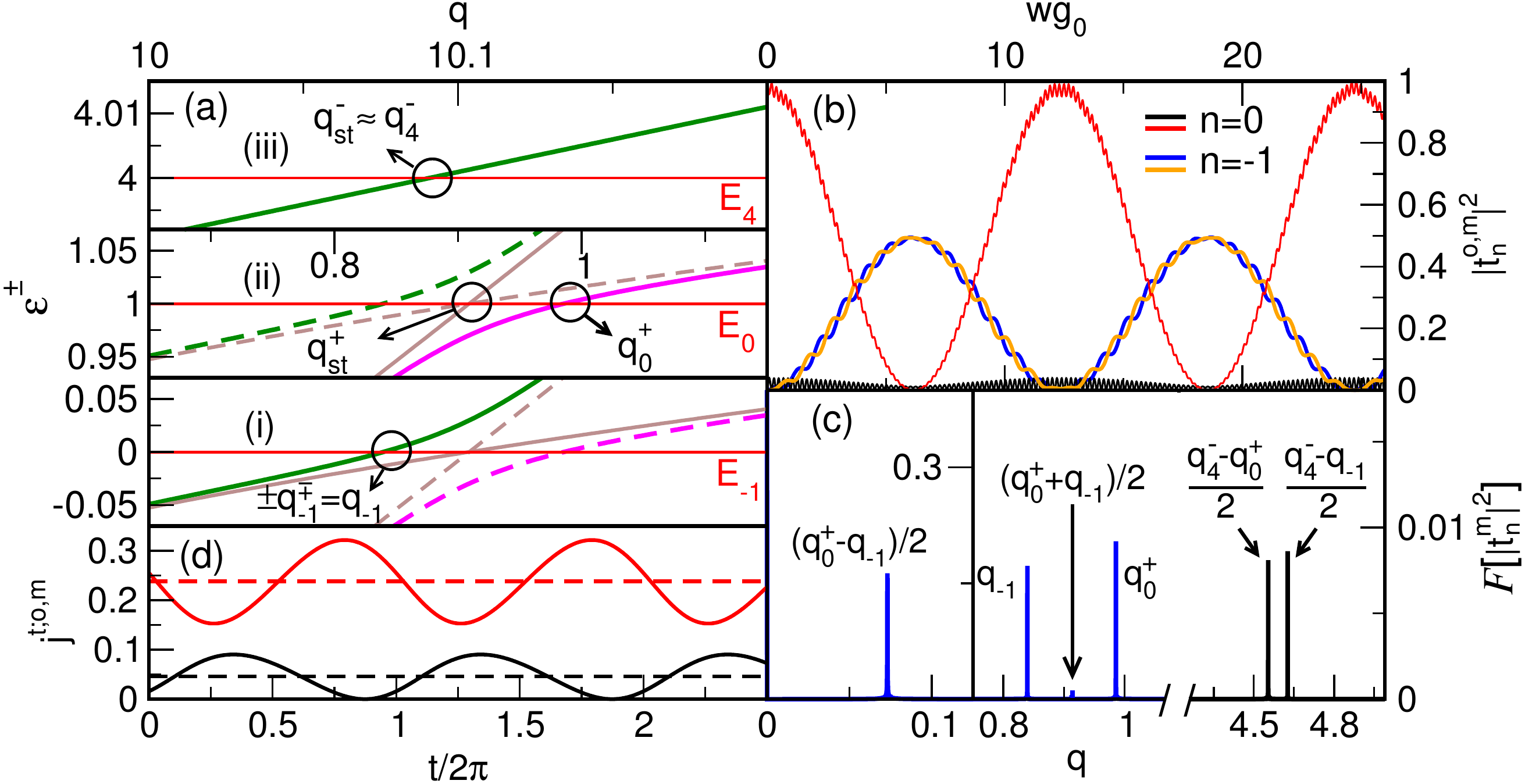}
\caption{(Color online)  
Scattering characteristics for photon energy $E=\Omega$. Notations are the same as in Fig.~\ref{fig3}. Note that in panel (a) for case (iii), due to avoided crossings, the wavenumber $q_{4}^{-}$ is obtained at $E_{4}$ (here, the static dispersion merges with the quasienergy). In panels (b) and (c), the blue lines correspond to the mechanical  sideband $n=-1$. In panel (d)  $wg_{0}= 0.929$ [crossing of black and blue lines in panel (b)]. Again, $g_{0}=0.287\Omega$, $g_{1}=0.073\Omega$ with $E=\Omega$ (corresponding to  $E/g_{0}=3.48$ in Fig.~\ref{fig1}).}\label{fig4}
\end{figure}
To elucidate the underlying mechanism, Fig.~\ref{fig2} shows the  quasienergies obtained from Floquet theory as a function of  $q$ for the parameters used in  Fig.~\ref{fig1}. Due to the optical and mechanical degrees of freedom, the quasienergies are two-fold degenerate. This leads to avoided crossings, appearing at energies $E=0$ and $E=\pm \Omega$ for the considered value of $g_{0}$, which are the reason for the  drastic modification of the transmission pattern  that becomes visible in the lower panels of Fig.~\ref{fig1}. The avoided crossings are displayed in Fig.~\ref{fig3} (a) in more detail, together with the energy dispersion for the static case. We observe that the oscillating barrier influences the scattering process the greater the wavenumbers $q_{0,1}$  deviate from the static wavenumbers $q_{\text{st}}$. The difference is largest in the vicinity of the avoided crossings. As a result, for an incoming photon with energy $E \simeq 0$, the entire transmission by  the optical and mechanical central bands is transferred to the first optical sidebands $n=\pm 1$ (periodically in  $wg_{0}$); see Fig.~\ref{fig3} (b) and  lower left panel in Fig.~\ref{fig1} at $E/g_{0}=10^{-3}$.  Since the situation is symmetric for the given parameter values, the wavenumbers obtained from the two quasienergies $\varepsilon^{\pm}$ have equal magnitudes but are antiparallel to each other,  cf. Fig.~\ref{fig3} (a).  Consequently, the interference of the central band and the sideband leads to standing optical and mechanical waves of different frequency. This becomes visible in the Fourier transform $F[\left|t_{0}^{m}\right|^2](q)$, see Fig.~\ref{fig3} (c). The interference effects are also reflected in the periodic time-evolution of the probability current density shown in Fig.~\ref{fig3} (d).  This is most strikingly demonstrated by  the photonic current (red solid line), which disappears periodically because of  destructive interference. The higher harmonic with frequency $2\Omega$ is due to the interference of the sidebands $n=\pm 1$.

A similar scenario arises for a photon at  resonance energy $E=\Omega$ (in Fig.~\ref{fig1} at $E/g_{0} \approx 3.48)$. Whereas for a static barrier, $g_{1}=0$, the transmission becomes almost purely photonic [cf.  Fig.~\ref{fig1}, upper right panel]  a small perturbation $g_{1} =0.073\Omega$ is sufficient to excite the sideband $n=-1$ [see Fig.~\ref{fig4}~(a)]. Again, interference of the central bands and the sidebands leads to periodic transmission probabilities as a function of $wg_{0}$, especially for the mechanical sideband, see Fig.~\ref{fig4} (b).  The Fourier transformation of the mechanical mode depicted in Fig.~\ref{fig4} (c) reveals which wavenumbers are involved in the scattering process. Just as for the photon current in Fig.~\ref{fig3} (d), the interference of  the mechanical side and central band leads to the suppression of the current density of the phonon periodically in time, see Fig.~\ref{fig4} (d).

{\it Conclusions.}
In optomechanical Dirac materials  scattering of plane photon waves by laser-induced oscillating planar barriers becomes inelastic.  Finite transmission probabilites for the optical and mechanical sidebands lead to a suppression or revivial of light-sound interconversion for photon energies close to multiple integers of the oscillation frequency. Using   parameter values of recent experiments~\cite{SPM15,SP10},  these effects will appear even for weak  couplings  with oscillation frequencies of about $0.5$MHz. Therefore, our work could be of particular interest for future (quantum) optical applications, especially for the experimental realization of an interface between microwave photons and phonons using laser barriers.   It would be desirable to extend this study to  more realistic quantum-dot geometries. This will be the subject of future work.
\vspace*{-0.3cm}


\end{document}